\pdfoutput=1
\documentclass[a4paper]{jpconf}

\usepackage{graphicx}

\usepackage{lmodern}
\usepackage[T1]{fontenc}
\usepackage{hyperref}

\newcommand\SCHOONSCHIP{{\tt SCHOONSCHIP}}
\newcommand\FORM{{\tt FORM}}
\newcommand\Forcer{{\tt Forcer}}
\newcommand\Mincer{{\tt Mincer}}
\newcommand\Python{{\tt Python}}
\newcommand\igraph{{\tt igraph}}

\newcommand{\reportnumber}[1]{%
  \makebox(0,0)[l]{%
    \parbox{\textwidth}{%
      \begin{flushright}%
        \normalsize #1
      \end{flushright}%
    }%
  }%
  \hskip 0pt plus 0.0001fil%
}

\begin{document}


\reportnumber{NIKHEF-2016-015}

\title{Calculating four-loop massless propagators \\ with \Forcer{}}

\author{T Ueda$^1$, B Ruijl$^{1,2}$ and J A M Vermaseren$^1$}

\address{$^1$ Nikhef Theory Group,
         Science Park~105, 1098~XG Amsterdam, The Netherlands}

\address{$^2$ Leiden University,
         Niels Bohrweg~1, 2333~CA Leiden, The Netherlands}

\ead{tueda@nikhef.nl, benrl@nikhef.nl, t68@nikhef.nl}

\begin{abstract}
We present \Forcer{}, a new \FORM{} program for the calculation of four-loop
massless propagators.
The basic framework is similar to that of the \Mincer{} program for three-loop
massless propagators:
the program reduces Feynman integrals to a set of master integrals in a
parametric way.
To overcome an ineludible complexity of the program structure at the four-loop
level, most of the code was automatically generated or made with
computer-assisted derivations.
Correctness of the program has been checked with the recomputation of some
quantities in the literature.
\end{abstract}


\section{Introduction}

Calculating massless propagator-type Feynman integrals is one of
the basic building
blocks in higher-order perturbative calculations in QCD
(see Ref.~\cite{Baikov:2015tea} for a recent review).
Up to the three-loop level, it turned out that dimensionally regularized~%
\cite{Bollini:1972ui,'tHooft:1972fi}
massless propagator-type
Feynman integrals can be reduced by the so-called triangle rule~%
\cite{Tkachov:1981wb,Chetyrkin:1981qh,Tkachov:1984xk}
obtained from integration-by-parts (IBP) identities~%
\cite{Tkachov:1981wb,Chetyrkin:1981qh}
and successive one-loop integrations with $G$-functions~%
\cite{Chetyrkin:1980pr,Chetyrkin:1981qh} except for two special topologies,
which are again reduced by manually solving IBP identities leading to two master
integrals~\cite{Chetyrkin:1981qh}.
Together with the results of the master integrals~\cite{Chetyrkin:1980pr},
this led to a program \Mincer{}~\cite{Gorishnii:1989gt},
implemented in \SCHOONSCHIP{}~\cite{Strubbe:1974vj} and later reprogrammed~%
\cite{Larin:1991fz} in \FORM{}~\cite{Kuipers:2012rf}.
\Mincer{} can analytically evaluate arbitrary scalar massless propagator-type
Feynman integrals up to the three-loop level as Laurent series expansions
in $\epsilon$, where $D=4-2\epsilon$ is the number of space-time dimensions.

On the other hand, at the four-loop level, people have used more systematic
or generic ways to perform the IBP reductions:
Laporta's algorithm~\cite{Laporta:2001dd} (for public implementations
see~\cite{%
Anastasiou:2004vj,%
Smirnov:2008iw,%
Smirnov:2013dia,%
Smirnov:2014hma,%
Studerus:2009ye,%
vonManteuffel:2012np%
}),
Baikov's method~\cite{Baikov:1996rk,Baikov:2005nv} and heuristic approaches~%
\cite{Lee:2012cn,Lee:2013mka}.
The values of the master integrals at the four-loop level are known
up to enough orders for practical applications~\cite{Baikov:2010hf,Lee:2011jt}.
Then one might think that the \Mincer{} program could be extended to
the four-loop
level, giving specialized (and hopefully optimized) routines for massless
propagator-type integrals, and this would be more efficient than
general-purpose IBP solvers.
This approach has not been pursued, however, to the best of our knowledge.
A reason may be that there are too many topologies at the four-loop
level, and therefore it is impractical to program treatment of all topologies
by hand as it was done for \Mincer{}.

The aim of this work is to develop a new \FORM{} program \Forcer{}~%
\cite{forcer}, which can be considered as an extension of \Mincer{} to
the four-loop level.
This requires automatized code generation because the large number of
topologies makes the program structure too complicated to program by hand.
Such a program is promising when one considers extremely time-consuming
computations with existing approaches, for example, higher moments of
four-loop splitting functions.

\newcommand\putgraphics[2][-0.5]{%
  \raisebox{#1\height}{\includegraphics{#2}}%
}
\begin{figure}
  \centering
  \renewcommand{\arraystretch}{1.4}
  \begin{tabular}{cc}
    \multicolumn{2}{c}{
      \putgraphics{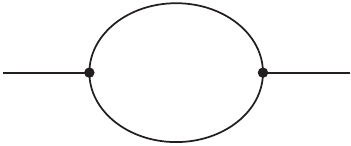}
      \hspace{5pt}
      \putgraphics{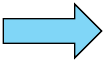}
      \hspace{5pt}
      \putgraphics{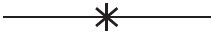}
    } \\
    \multicolumn{2}{c}{(a) one-loop insertion} \\[10pt]
    \multicolumn{2}{c}{
      \putgraphics[-0.4]{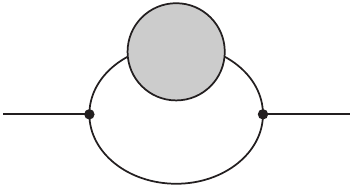}
      \hspace{5pt}
      \putgraphics{fig-rightarrow.pdf}
      \hspace{5pt}
      \putgraphics{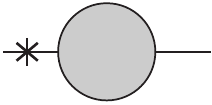}
    } \\
    \multicolumn{2}{c}{(b) one-loop carpet} \\[10pt]
    \putgraphics{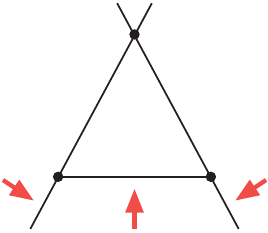} {} &
    \putgraphics{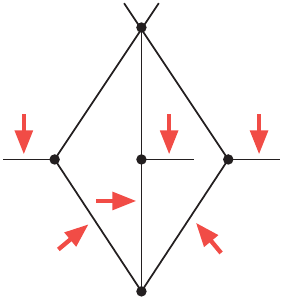} \\
    (c) triangle & (d) diamond
  \end{tabular}
  \caption{%
    Substructures of topologies that lead to reduction to simpler topologies.
    (a) One-loop insertion.
    A line with a non-integer power is indicated with ``$\ast$''.
    (b) One-loop carpet.
    (c) Triangle.
    The triangle rule removes one of the three lines indicated by arrows.
    (d) Two-loop diamond.
    The diamond rule removes one of the six lines indicated by arrows.
  }
  \label{fig1}
\end{figure}

\section{Reduction for each topology}

Here we briefly summarize how one can reduce integrals belonging to a topology
into those in simpler topologies from a diagrammatical point of view.
Techniques of \ref{2.1}, \ref{2.2} and \ref{2.3} were used in
the three-loop \Mincer{} program.
The technique of \ref{2.4} is new at the four-loop level.

\subsection{One-loop insertion integral}
\label{2.1}

When a topology contains a massless one-loop insertion graph as
its substructure and both propagators are massless, one can
integrate out the corresponding loop momentum and it becomes a line but
gets a non-integer power $1/(Q^2)^\epsilon$ in the resultant momentum $Q$
(figure~\ref{fig1}a).
The powers of the two propagators can be arbitrary numbers.
It is also allowed to have polynomic tensorial numerators with respect to the
loop momentum.

\subsection{One-loop carpet integral}
\label{2.2}

This is another type of massless one-loop integrals that we can perform.
When a subgraph is inserted to a one-loop graph, the outer loop can be
integrated out first (figure~\ref{fig1}b).
Similar to the one-loop insertion integrals, the powers of the two propagators
of the outer loop can be arbitrary numbers.
Numerators are also allowed.

\subsection{Triangle rule}
\label{2.3}

When a topology contains a one-loop triangle graph depicted in
figure~\ref{fig1}c as a subgraph, recursive usage of an IBP identity leads to
a sum of integrals corresponding to simpler topologies with one of the three
lines indicated by arrows removed. This is possible when the powers of
the three
propagators are integer; hence this rule cannot be applied if
any of the three lines has taken non-integer powers from one-loop integrals
explained above. Irreducible numerators in the topology can always be chosen
such that they do not interfere with the rule.

\subsection{Diamond rule}
\label{2.4}

The triangle rule can be extended for multi-loop diamond-shaped subgraphs~%
\cite{Ruijl:2015aca}.
Figure~\ref{fig1}d shows the simplest diamond structure, in which one of six
lines is removed in the end of the recursion.

\subsection{Manual rules}

If none of the above substructures is available in a topology,
one needs to manually solve IBP
identities such that recursive use of rules removes one of the propagators,
or at least simplify integrals as much as possible.
In the latter case, irreducible integrals are considered as master integrals
of the problem.
This part has not been fully automatized yet, though we used computer-assisted
derivations of reduction rules.

Starting from a set of IBP identities obtained in a normal way, one can generate
sets of identities by raising or lowering one of (or combinations of) powers of
propagators and irreducible numerators in all possible ways.
They are combined into a new set of identities, and then complicated integrals
are eliminated by means of Gaussian elimination.
Powers of the propagators remain parametric.
We construct an reduction scheme out of the resulted identities.

\section{Code generation for all topologies}

In order to handle many topologies and transitions among them
at the four-loop level,
we need to automatize code generation for them.

First, we represent a topology as an undirected graph in graph theory.
This makes it easy to detect the substructures explained in the previous
section by pattern matchings of connections of vertices and edges.
We implemented such routines in \Python{} with a graph library
\igraph{}~\cite{igraph}.

Then, starting from the top-level topologies, we consider to remove a line
from each topology in all possible ways.
If removing a line gives a massless tadpole, then the generated topology
is immediately discarded because it is zero.
Some of the generated topologies are actually identical and
such graph isomorphisms
are efficiently detected by the graph library.
We keep track of all mappings of momenta between topologies before and after
the transition. They are needed to rewrite propagators and irreducible
numerators in a topology into those in another topology.

For each topology, the next action is determined from its substructures.
Irreducible numerators are chosen such that they do not interfere with
the next action.
An adequate subroutine must be called for the next action in generated code.
Symmetries in each topology are detected as graph automorphisms, which are
helpful to reduce the number of terms we need to process in actual calculations.

\begin{figure}
  \centering
  \includegraphics[width=\textwidth]{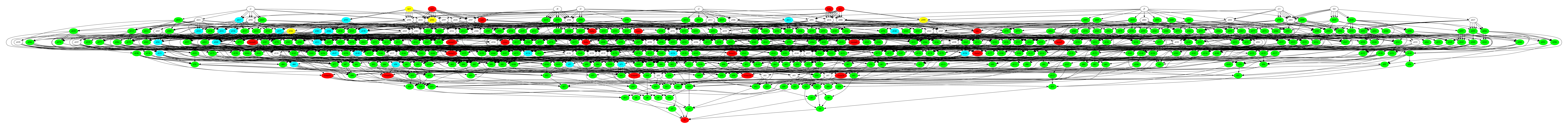}
  \begin{picture}(0,0)%
    \setlength{\unitlength}{1cm}
    \linethickness{0.05cm}
    \put(-0.08,0.35){\dashbox{0.15}(1.65,1.50){}}
  \end{picture}
  \includegraphics[trim=5050 0 4100 0,clip,width=\textwidth]{reductionflow4.pdf}
  \caption{%
    The tree-like graph structure of topology transitions at the four-loop
    level.
    The lower figure shows the enlarged view of the boxed region in the upper
    one.
    Green, cyan, white, yellow and red circles represent topologies in which
    one-loop insertion integral, carpet integral, triangle rule, diamond rule
    and none of them are available, respectively.
  }
  \label{fig2}
\end{figure}

Repeating this procedure until all topologies are reduced into a trivial
Born graph,
we obtain a tree of all possible topologies (figure~\ref{fig2}).
From 11 top-level topologies at the four-loop level, we obtained 437 non-trivial
topologies in total.
The numbers of topologies that have one-loop insertions, one-loop carpets,
triangles and diamonds as their substructures are 335, 24, 53 and 4,
respectively, and
21 topologies need construction of manual rules.
From the topology tree, we generated \FORM{} code for reductions in
all topologies, and rewriting propagators and irreducible numerators
at all topology transitions.

\section{Conclusion}

We have constructed a \FORM{} program \Forcer{}, which analytically computes
massless propagator-type Feynman integrals up to the four-loop level.
The complicated program structure made us opt to go for automatization of
code generation.
The problem that there are many topologies at this level was solved by
representing all topology transitions as a topology tree and generate
code in an automatized way.
It is difficult to automatize deriving manual rules for topologies
where none of known substructures is available.
However, we observed that it is
helpful to employ computer-assisted derivations and especially
Gaussian elimination with a ordering of integrals by their complexities,
even for parametric reduction rules.

In order to check correctness of the program, we performed recomputations of
several known results in the literature.
We recomputed the four-loop QCD $\beta$-function~%
\cite{vanRitbergen:1997va,Czakon:2004bu} and checked the gauge-invariance of
the result.
Moreover, we computed the same quantity in the background field method and
obtained the same result.
Another set of recomputations were done for four-loop
anomalous dimensions of fixed
$N$-moments of the non-singlet twist-2 operator. We reproduced
$N=2$~\cite{Baikov:2006ai,Velizhanin:2011es}, $N=3$ and $N=4$ results~%
\cite{Velizhanin:2014fua,sfb}.
More physics results obtained by \Forcer{} will be reported elsewhere~%
\cite{forcer-physics}.


\ack

We would like to thank Andreas Vogt for discussions and collaboration
for physics applications.
This work is supported by the ERC Advanced Grant no. 320651, ``HEPGAME''.

\section*{References}

\bibliography{mybibfile}

\end{document}